\documentclass[aps,prl,twocolumn,groupedaddress,showpacs]{revtex4}%
\usepackage{graphicx}
\usepackage{amsmath}
\usepackage{dcolumn}
\usepackage{bm}
\usepackage{amsfonts}
\usepackage{amssymb}%
\ifx\pdfoutput\relax\let\pdfoutput=\undefined\fi
\newcount\msipdfoutput
\ifx\pdfoutput\undefined\else
\ifcase\pdfoutput\else
\ifx\paperwidth\undefined\else
\ifdim\paperheight=0pt\relax\else\pdfpageheight\paperheight\fi
\ifdim\paperwidth=0pt\relax\else\pdfpagewidth\paperwidth\fi
\fi\fi\fi
\begin{document}
\title{Spatial Orientation using Quantum Telepathy}
\author{F.A. Bovino}
\email{fabio.bovino@elsag.it}
\affiliation{Elsag spa, Via Puccini 2, 16154 Genova, Italy}
\author{M. Giardina }
\affiliation{Elsag spa, Via Puccini 2, 16154 Genova, Italy}
\author{K. Svozil}
\affiliation{Institut f\"{u}r Theoretische Physik, University of Technology Vienna,Wiedner
Hauptstra\ss e 8-10/136, A-1040 Vienna, Austria}
\author{V. Vedral}
\affiliation{Quantum Information Technology Lab, Department of Physics, National University
of Singapore, Singapore 117542}
\affiliation{The School of Physics and Astronomy, University of Leeds, Leeds LS2 9JT, UK}
\date{\today}

\begin{abstract}
We implemented the protocol of entanglement assisted orientation in the space
proposed by Brukner et al. (quant-ph/0509123). We used min-max principle to
evaluate the optimal entangled state and the optimal direction of polarization
measurements which violate the classical bound.

\end{abstract}

\maketitle






Bizarre effects of quantum entanglement \cite{Schrodinger35},\cite{EPR35}, are
usually dramatized using Bell's inequalities \cite{Bell64},\cite{CHSH69}%
,\cite{FC72},\cite{Tsirelson80}. These show that correlations between
measurements on two spatially separated systems can be higher than anything
allowed by the "local realistic" (i.e. classical) theories. The way that
testing Bell's inequalities almost invariably proceeds is, in very broad
terms, as follows. Alice and Bob share a number of entangled pairs, and Alice
measures her systems at the same time as Bob measures his systems. After that,
they communicated classically their results to each other and compute various
correlation functions. When they combine these correlation functions into a
Bell's inequality, they can then check if the inequality is violated
(signifying the existence of correlations stronger than any classical one). It
is crucial for this experiment that Alice and Bob classically communicate with
each other. Otherwise they would never be able to compute the necessary
correlation functions in order to test the inequality. It is absolutely
extraordinary, however, that there are applications where Alice and Bob could
utilize stronger than classical correlations without any form of classical
communication. \begin{figure}[h]
\includegraphics[angle=0,width=7.5truecm]{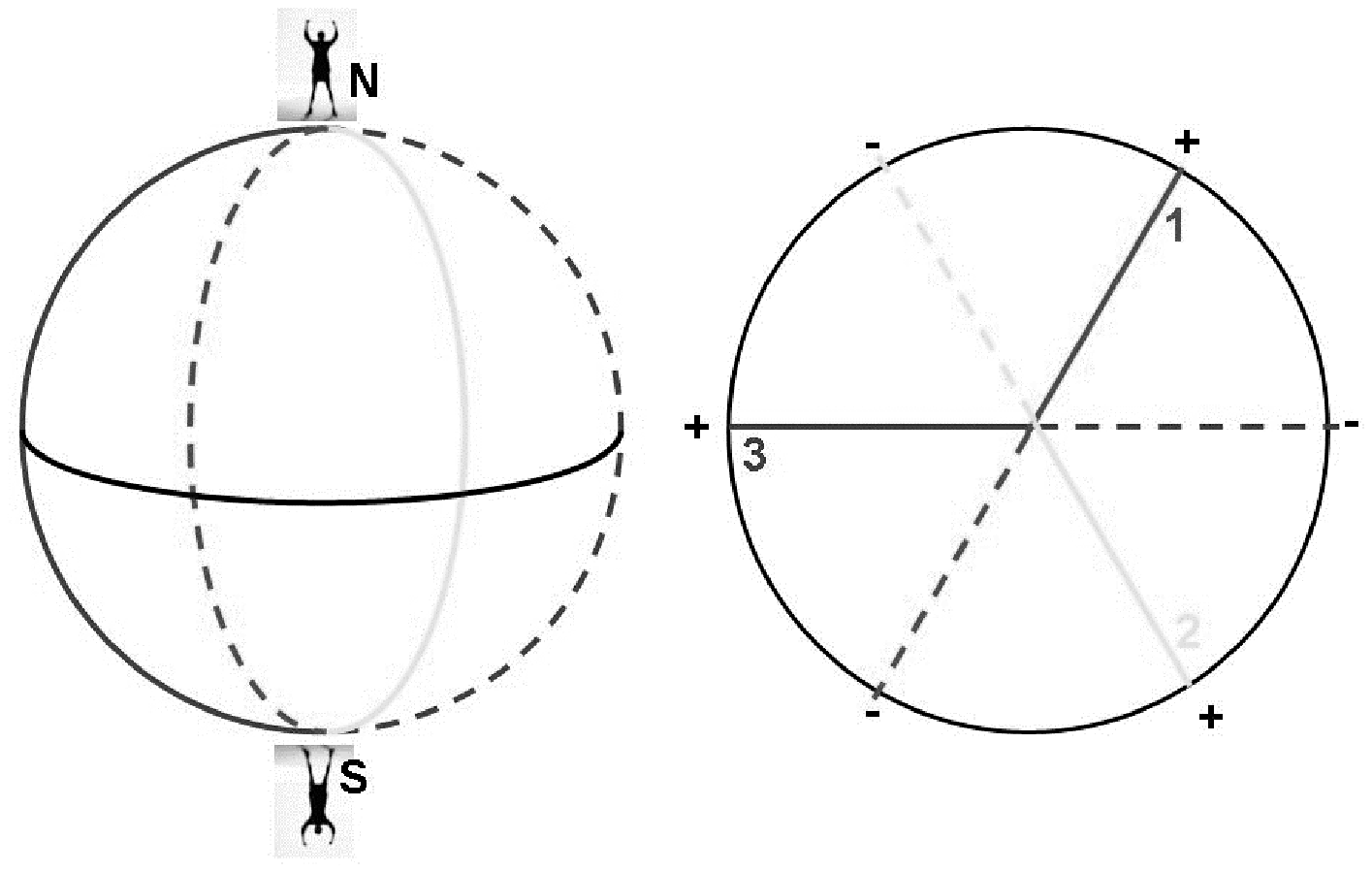}\caption{Two partners
(Alice and Bob) are on the two poles of the Earth: there are three paths and
two directions (+ and -) for each path: each partner have to find the other in
the lack of any classical communication. To achieve their goal the best
strategy is to maximize the probability to take the same directions, if they
choose the same path, and the probability to take opposite directions if they
choose different paths.}%
\label{protocol}%
\end{figure}Suppose that Alice and Bob are far away from each other, but
happen to share some entanglement (this could have been established when they
met at some earlier time). Can they, using entanglement but without utilizing
any classical communication, move in the direction towards each other faster
than allowed by any local realistic theories? Namely can they find each other
without communication? Surprisingly, this protocol is possible as shown very
recently by Brukner et al in \cite{Brukner05}. The way that this would proceed
is that, depending on the outcomes of their respective measurements, Alice and
Bob would move in certain directions, and entanglement would ensure that the
directions are such that they (on average) approach each other faster than
allowed classically and yet without communicating with each other. This
protocol clearly exemplifies why entanglement deserves to be called "spooky".
The effect could, in fact, be called "spatial orientation using quantum telepathy".

In this letter we experimentally demonstrate that quantum entanglement indeed
leads to the faster than classical orientation in space. \begin{figure}[h]
\includegraphics[angle=0,width=7.5truecm]{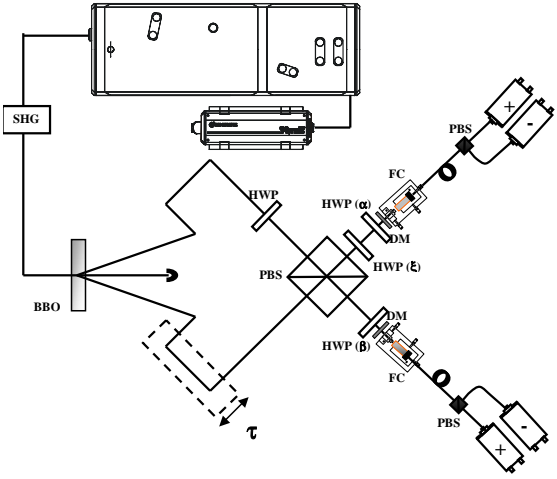}\caption{Experimental
set-up. A 3\thinspace mm long $\beta$-barium borate crystal, cut for a Type.II
phase-matching, is pumped in ultrafast regime.The SPDC photon pairs, are
generated as coherent superposition of $\left\vert HV\right\rangle $ and
$\left\vert VH\right\rangle $. The HWP changes the two alternatives in
$\left\vert HH\right\rangle $ and $\left\vert VV\right\rangle $.The PBS
provides the symmetrization of amplitude probabilities. The temporal
superposition of the two alternatives is reached by changing the length of the
trombone $\left(  \tau\right)  $. At the output of the interferometer the Bell
state $\left\vert \Phi^{+}\right\rangle $ is synthesized. By tilting the BBO
crystal and rotating the third HWP it is possible to synthesize all Bell
States or a linear combinationof two of them \cite{KKCGS03},\cite{BCDC04}. }%
\label{setup}%
\end{figure}Two partners (Alice and Bob) are on the two poles of the Earth;
there are three paths and two directions (+ and -) for each path: each partner
have to find the other in the lack of any classical communication
(Fig.\ref{protocol}). To achieve their goal the best strategy is to maximize
the probability to take the same directions, if they choose the same path, and
the probability to take opposite directions if they choose different paths.
The overall probability of success is given by%
\begin{equation}
P=\frac{1}{9}\left(  \sum\limits_{i=1}^{3}P_{ii}\left(  same\right)
+\sum\limits_{i\neq j=1}^{3}P_{ij}\left(  opp\right)  \right)
\end{equation}
where $P_{ij}\left(  opp\right)  $ is the probability that Alice and Bob take
opposite direction, if they choose different paths, $P_{ii}\left(
same\right)  $ is the probability that they take the same direction if they
choose the same path.

The probability of success of any classical protocol is bounded by the value
7/9, because it was demonstrated that%
\begin{equation}
\beta=\sum\limits_{i=1}^{3}P_{ii}\left(  same\right)  +\sum\limits_{i\neq
j=1}^{3}P_{ij}\left(  opp\right)  \leq7 \label{beta}%
\end{equation}
holds for all local realistic models \cite{Brukner05}.

To increase the probability of success, Alice and Bob can share
polarization-entangled photon pairs: every partner independently choose a path
at random from the set \{1,2,3\}. The choice of the path determines a choice
of direction of polarization measurements: the possible outputs (+ or -) fix
the direction along the path.

The aim of this letter is to use the min-max principle to evaluate the optimal
entangled state and the optimal direction of polarization measurements which
violate the classical bound.

The min-max principle for self-adjoint transformations \cite{Halmos74} states
that the operator norm is bounded by the minimal and maximal eigenvalues. The
norm of the self-adjoint transformation resulting from the sum of the quantum
counterparts of all the classical terms contributing to a particular Bell
inequality obeys the min-max principle. Thus determining the maximal
violations of classical Bell inequalities amounts to solving an eigenvalue
problem. The associated eigenstates are the multi-partite states which yield a
maximum violation of the classical bounds under the given experimental setup
\cite{Werner01},\cite{Filipp04},\cite{Cabello04}.

In order to evaluate the quantum counterpart of the inequality (\ref{beta}),
the classical probabilities have to be substituted by the quantum ones. Let us
consider a two spin 1/2 particles configuration, described by its density
matrix $\rho$, in which the two particles move in opposite directions along
the y axis and the spin components are measured in the x-z plane. In such a
case, the single particle spin-up and down observables along $\vartheta_{i}$,
$\vartheta_{j}$, correspond to the projections $A_{\pm}\left(  \vartheta
_{i}\right)  $, with%
\begin{equation}
A_{\pm}\left(  \vartheta\right)  =\frac{1}{2}\left(  \mathbf{I}\pm
\mathbf{n}\left(  \vartheta\right)  \mathbf{\sigma}\right)
\end{equation}
where $\mathbf{\sigma}$ is the vector of the Pauli matrices. The joint
probability $q_{ij\text{ }}$for finding the left particle in the spin-up state
along the angle $\vartheta_{i}$ and the right particle in the spin-up state
along the angle $\vartheta_{j}$ is given by%
\begin{equation}
q_{ij\text{ }}=tr\{\rho\lbrack A_{+}\left(  \vartheta_{i}\right)  \otimes
A_{+}\left(  \vartheta_{j}\right)  ]\}. \label{correlations}%
\end{equation}
Then, substituting in the inequality (\ref{beta}), we obtain%
\begin{align}
P_{ii}\left(  same\right)   &  =tr\{\rho\lbrack A_{+}\left(  \vartheta
_{i}\right)  \otimes A_{+}\left(  \vartheta_{i}\right) \nonumber\\
&  +A_{-}\left(  \vartheta_{i}\right)  \otimes A_{-}\left(  \vartheta
_{i}\right)  ]\},\nonumber\\
P_{ij}\left(  opp\right)   &  =tr\{\rho\lbrack A_{+}\left(  \vartheta
_{i}\right)  \otimes A_{-}\left(  \vartheta_{j}\right) \nonumber\\
&  +A_{-}\left(  \vartheta_{i}\right)  \otimes A_{+}\left(  \vartheta
_{j}\right)  ]\}.
\end{align}
We are interested in maximal violations of the inequality (\ref{beta}) with
three possible measurements setting per observer: Alice and Bob choose between
three dichotomic observables, determined by three measurements angles
$\vartheta_{1}$, $\vartheta_{2}$, $\vartheta_{3}$. \begin{figure}[h]
\includegraphics[angle=0,width=8truecm]{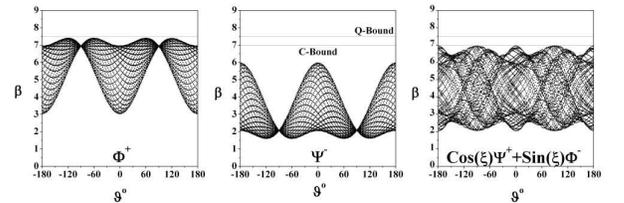}\caption{Experimentally
reconstructed bounds for eigenvalues $\lambda_{1,2,3,4}\left(  \varphi
,\vartheta\right)  $. The bounds for eigenvalues $\lambda_{3,4}\left(
\varphi,\vartheta\right)  $ were reached by the linear combination of the Bell
states $\left\vert \Psi^{+}\right\rangle $ and $\left\vert \Phi^{-}%
\right\rangle $. }%
\label{bound}%
\end{figure}The first possible choice is given by $\vartheta_{1}=0$,
$\vartheta_{2}=2\varphi$, $\vartheta_{3}=2\vartheta$. In this case the
eigenvalues $\lambda_{1,2,3,4}$, and the eigenvectors $\nu_{1,2,3,4}$,
corresponding to the maximal violating eigenstates of the self-adjoint
operator $O_{33}$%
\begin{align}
O_{33}  &  =\sum\limits_{s\in\{+,-\}}\sum\limits_{i=1}^{3}A_{s}\left(
\vartheta_{i}\right)  \otimes A_{s}\left(  \vartheta_{i}\right) \nonumber\\
&  +\sum\limits_{s\neq t\in\{+,-\}}\sum\limits_{i\neq j=1}^{3}A_{s}\left(
\vartheta_{i}\right)  \otimes A_{t}\left(  \vartheta_{j}\right)  \label{pippo}%
\end{align}
are%
\begin{align}
\lambda_{1}\left(  \varphi,\vartheta\right)   &  =6-\cos\left(  2\vartheta
\right)  -\cos\left(  2(\vartheta-\varphi)\right)  -\cos\left(  2\varphi
\right) \nonumber\\
\lambda_{2}\left(  \varphi,\vartheta\right)   &  =3+\cos\left(  2\vartheta
\right)  +\cos\left(  2(\vartheta-\varphi)\right)  +\cos\left(  2\varphi
\right) \nonumber\\
\lambda_{3}\left(  \varphi,\vartheta\right)   &  =\frac{1}{2}\left\{
9-\left[  15+2\cos\left(  4\vartheta\right)  -4\cos\left(  2(\vartheta
-2\varphi)\right)  \right.  \right. \nonumber\\
&  -4\cos\left(  2(2\vartheta-\varphi)\right)  +2\cos\left(  4(\vartheta
-\varphi)\right) \nonumber\\
&  \left.  \left.  +2\cos\left(  4\varphi\right)  -4\cos\left(  2(\vartheta
+\varphi)\right)  \right]  ^{1/2}\right\} \nonumber\\
\lambda_{4}\left(  \varphi,\vartheta\right)   &  =\frac{1}{2}\left\{
9+\left[  15+2\cos\left(  4\vartheta\right)  -4\cos\left(  2(\vartheta
-2\varphi)\right)  \right.  \right. \nonumber\\
&  -4\cos\left(  2(2\vartheta-\varphi)\right)  +2\cos\left(  4(\vartheta
-\varphi)\right) \nonumber\\
&  \left.  \left.  +2\cos\left(  4\varphi\right)  -4\cos\left(  2(\vartheta
+\varphi)\right)  \right]  ^{1/2}\right\}
\end{align}%
\begin{align}
\nu_{1}  &  =\left\vert \Phi^{+}\right\rangle \nonumber\\
\nu_{2}  &  =\left\vert \Psi^{-}\right\rangle \nonumber\\
\nu_{3}  &  =F\left(  \varphi,\vartheta\right)  \left\vert \Psi^{+}%
\right\rangle +G\left(  \varphi,\vartheta\right)  \left\vert \Phi
^{-}\right\rangle \nonumber\\
\nu_{4}  &  =H\left(  \varphi,\vartheta\right)  \left\vert \Psi^{+}%
\right\rangle +I\left(  \varphi,\vartheta\right)  \left\vert \Phi
^{-}\right\rangle
\end{align}
where the eigenvectors $\nu_{3}$ and $\nu_{4}$ are given by the superposition
of the Bell's states $\left\vert \Psi^{+}\right\rangle $, $\left\vert \Phi
^{-}\right\rangle $, by the functions F, G, H, I. The maximum eigenvalue is
$\lambda_{1}\left(  \varphi,\vartheta\right)  $ with the corresponding
eigenvector $\left\vert \Phi^{+}\right\rangle $, and optimal angles of
measurement given by $\left(  60^{o},-60^{o}\right)  $ (when we consider
angles less than $90^{o}$), where is achieved the value 7.5. The second
eigenvalue $\lambda_{2}\left(  \varphi,\vartheta\right)  $ with eigenvector
$\left\vert \Psi^{-}\right\rangle $, determinates the minimum bound for the
inequality (\ref{beta}). For the angles $\left(  60^{o},-60^{o}\right)  $, the
minimum value 1.5 is reached. The eigenvalues $\lambda_{3}\left(
\varphi,\vartheta\right)  $ and $\lambda_{4}\left(  \varphi,\vartheta\right)
$ stay always under the classical bound $7$. For a single value
parametrization, for example, $\vartheta_{1}=0$, $\vartheta_{2}=2\vartheta$,
$\vartheta_{3}=-2\vartheta,$ the eigenvalues $\lambda_{1,2,3,4}$, and the
eigenvectors $\nu_{1,2,3,4}$, corresponding to the maximal violating
eigenstates of the self-adjoint operator $O_{33}$ are%
\begin{align}
\lambda_{1}  &  =6-2\cos\left(  2\vartheta\right)  -\cos\left(  4\vartheta
\right)  \text{, }\nu_{1}=\left\vert \Phi^{+}\right\rangle \nonumber\\
\lambda_{2}  &  =5+2\cos\left(  2\vartheta\right)  -\cos\left(  4\vartheta
\right)  \text{, }\nu_{2}=\left\vert \Psi^{+}\right\rangle \nonumber\\
\lambda_{3}  &  =4-2\cos\left(  2\vartheta\right)  +\cos\left(  4\vartheta
\right)  \text{, }\nu_{3}=\left\vert \Phi^{-}\right\rangle \nonumber\\
\lambda_{4}  &  =3+2\cos\left(  2\vartheta\right)  +\cos\left(  4\vartheta
\right)  \text{, }\nu_{4}=\left\vert \Psi^{-}\right\rangle
\end{align}
then the entangled state $\left\vert \Phi^{+}\right\rangle $ provides the
violation of classical bound for $\vartheta=60^{o}$. In this case to any
eigenvalue one Bell state corresponds. \begin{figure}[hh]
\includegraphics[angle=0,width=7.5truecm]{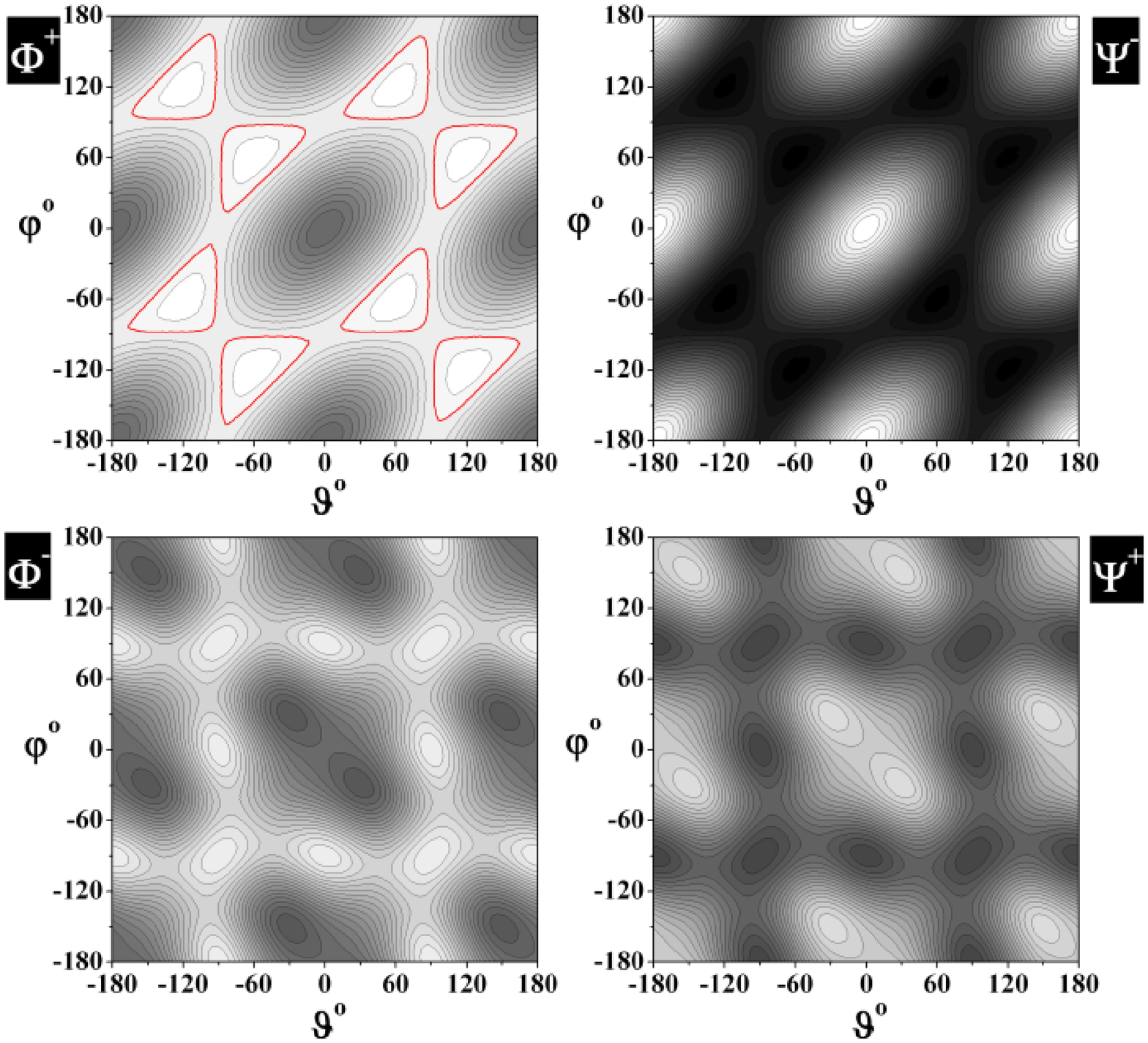}\caption{The contour
plots represent the experimental reconstruction of $\beta$ for the
bidimesional parametrization $\left(  0,2\varphi,2\vartheta\right)  $. Only
the state $\left\vert \Phi^{+}\right\rangle $ violates the classical bound 7.
For the states $\left\vert \Phi^{+}\right\rangle $ and $\left\vert \Psi
^{-}\right\rangle $ the corresponding plots represents, also, the
bidimensional eigenvalues $\lambda_{1,2}\left(  \varphi,\vartheta\right)  $,
i.e., the maximum and the minimum bound of $O_{33}$. The state $\left\vert
\Phi^{-}\right\rangle $ reaches the classical bound 7.}%
\label{newbell}%
\end{figure}
In the experimental set-up (see Fig.\ref{setup}), a 3\thinspace mm long
$\beta$-barium borate crystal, cut for a TypeII phase-matching
\cite{Klyshko88}, \cite{KMWZSS95}, \cite{Rubin96}, is pumped in ultrafast
regime (120\thinspace fs) by a train of $\Omega_{\mathit{pump}}=410$\thinspace
nm pulses generated by the second harmonic of a Ti:Sa laser. SPDC (Spontaneous
Parametric Down-Converted) photon pairs at 820\thinspace nm $\left(
\Omega_{\mathit{pump}}/2\right)  $ are generated with an emission angle of
$3^{o}$. After passing through the interferometer, thanks to temporal
engineering and amplitude symmetrization, we obtain the entangled state%
\begin{equation}
\left\vert \Phi^{+}\right\rangle =\frac{1}{\sqrt{2}}\left(  \left\vert
HH\right\rangle +\left\vert VV\right\rangle \right)
\end{equation}
where $\emph{H}$ ($\emph{V}$) stays for Horizontal (Vertical). The photons are
coupled by lenses into single-mode fibers. Coupling efficiency has been
optimized by a proper engineering of the pump and the collecting mode in
experimental conditions \cite{BVCCDS03}. Dichroic mirrors are placed in front
of the fiber couplers to reduce stray light due to pump scattering. Half Wave
Plates (HWPs) before the fiber coupler, together with fiber-integrated
polarizing beam splitters (PBSs), project photons in the polarization basis
$\left\vert s\left(  2\vartheta\right)  \right\rangle =\cos\left(
\vartheta\right)  \left\vert H\right\rangle +\sin\left(  \vartheta\right)
\left\vert V\right\rangle $, $\left\vert s^{\bot}\left(  2\vartheta\right)
\right\rangle =\sin\left(  \vartheta\right)  \left\vert H\right\rangle
-\cos\left(  \vartheta\right)  \left\vert V\right\rangle $. Photons are
detected by single photon counters (Perkin-Elmer SPCM-AQR-14). A third HWP
($\xi$) provides to prepare the superposition of two Bell states ($\left\vert
\Psi^{+}\right\rangle $ and $\left\vert \Phi^{-}\right\rangle $) to
experimentally reconstruct all two-dimensional bounds \cite{BCDC04}.

The local observables $\hat{A}_{\pm}(\vartheta_{i})$ can be rewritten for the
chosen polarization basis $\left\{  \left\vert s(2\vartheta)\right\rangle
,\left\vert s^{\bot}(2\vartheta)\right\rangle \right\}  $ as%

\begin{align}
\hat{A}_{+}  &  =|s(2\vartheta)\rangle\langle s(2\vartheta)|\nonumber\\
\hat{A}_{-}  &  =|s^{\bot}(2\vartheta)\rangle\langle s^{\bot}(2\vartheta)|
\end{align}

and the correlation functions (\ref{correlations}) can be expressed in terms
of coincidence detection probabilities $p_{x,y}\left(  \vartheta_{i}%
,\vartheta_{j}\right)  $ as:%
\begin{align}
&  \left\langle A_{+}\left(  \vartheta_{i}\right)  \otimes A_{+}\left(
\vartheta_{i}\right)  +A_{-}\left(  \vartheta_{i}\right)  \otimes A_{-}\left(
\vartheta_{i}\right)  \right\rangle \nonumber\\
&  =p_{++}\left(  \vartheta_{i},\vartheta_{i}\right)  +p_{--}\left(
\vartheta_{i},\vartheta_{i}\right) \\
&  \left\langle A_{+}\left(  \vartheta_{i}\right)  \otimes A_{-}\left(
\vartheta_{j}\right)  +A_{-}\left(  \vartheta_{i}\right)  \otimes A_{+}\left(
\vartheta_{j}\right)  \right\rangle \nonumber\\
&  =p_{+-}\left(  \vartheta_{i},\vartheta_{j}\right)  +p_{-+}\left(
\vartheta_{i},\vartheta_{j}\right)
\end{align}
where $x,y=+,-$ are the two outputs of the integrated PBS and $p_{x,y}\left(
\vartheta_{i},\vartheta_{j}\right)  $ are expressed in terms of coincident
counts:%
\begin{equation}
p_{x,y}\left(  \vartheta_{i},\vartheta_{j}\right)  =\frac{N_{x,y}\left(
\vartheta_{i},\vartheta_{j}\right)  }{N_{TOT}}%
\end{equation}
where $N_{x,y}\left(  \vartheta_{i},\vartheta_{j}\right)  $ is the number of
coincidences measured by the pair of detectors $x,y$ in the above described
polarization basis, and $N_{TOT}=N_{++}\left(  \vartheta_{i},\vartheta
_{j}\right)  +N_{+-}\left(  \vartheta_{i},\vartheta_{j}\right)  +N_{-+}\left(
\vartheta_{i},\vartheta_{j}\right)  +N_{--}\left(  \vartheta_{i},\vartheta
_{j}\right)  $. \begin{figure}[h]
\includegraphics[angle=0,width=7.5truecm]{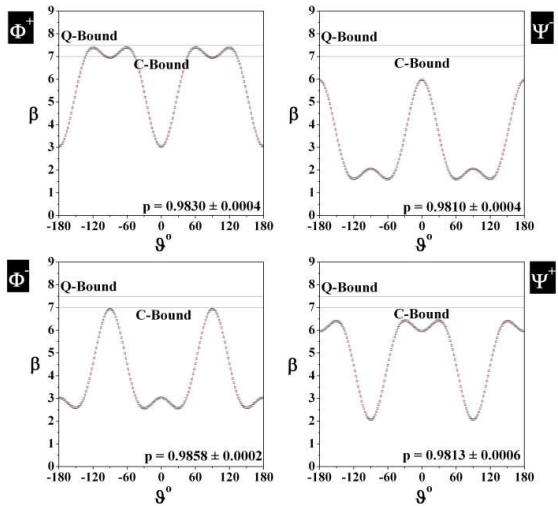}\caption{Experimental
reconstruction of $\beta$ for parametrization $\left(  0,2\vartheta
,-2\vartheta\right)  $. A non-ideal state affected by white noise can be
written as: $p\left\vert \Phi^{+}\right\rangle \left\langle \Phi
^{+}\right\vert +\frac{\left(  1-p\right)  }{4}I.$ The maximum experimental
value is $\beta\simeq7.41$ and from the corresponding fit function we obtained
the value $p\simeq0.98$.}%
\label{meigen}%
\end{figure}In Fig.\ref{bound} we show the experimentally reconstructed bounds
for eigenvalues $\lambda_{1,2,3,4}\left(  \varphi,\vartheta\right)  $. The
bounds for eigenvalues $\lambda_{3,4}\left(  \varphi,\vartheta\right)  $ are
reached by the linear combination of the Bell states $\left\vert \Psi
^{+}\right\rangle $ and $\left\vert \Phi^{-}\right\rangle $. In addition, in
Fig.\ref{newbell} we show the contour plots representing the experimental
reconstruction of the Bell operator $\beta$ for the bi-dimensional
parametrization $\left(  0,2\varphi,2\vartheta\right)  $: only the state
$\left\vert \Phi^{+}\right\rangle $ violates the classical bound $7$. For the
states $\left\vert \Phi^{+}\right\rangle $ and $\left\vert \Psi^{-}%
\right\rangle $ the corresponding plots represents, also, the bi-dimensional
eigenvalues $\lambda_{1,2}\left(  \varphi,\vartheta\right)  $, i.e., the
maximum and the minimum bound of $O_{33}$. The state $\left\vert \Phi
^{-}\right\rangle $ reaches the classical bound $7$. In Fig.\ref{meigen} we
show the experimental reconstruction of $\beta$ for the mono-dimensional
parametrization $\left(  0,2\vartheta,-2\vartheta\right)  $, and, in
particular, the violation of the maximum values of the Bell operator $\beta$
for the state $\left\vert \Phi^{+}\right\rangle .$ Due to the experimental
imperfections (misalignment and presence of stray light),\ the state generated
from the source could be written as $p\left\vert \Phi^{+}\right\rangle
\left\langle \Phi^{+}\right\vert +\frac{\left(  1-p\right)  }{4}I$, including
a white noise term: from the experimental value $\beta\simeq7.41$ and the
corresponding fit function, we obtained $p\simeq0.98$.

Thus it could seem not surprising that a maximally entangled state is the one
violating classical forecasts and providing a "speed-up" in spatial
orientation, the actual demonstration of this conclusion is not obvious and
could be not valid for different Bell's like inequalities. Moreover, the fact
that the $\left\vert \Phi^{+}\right\rangle $ state, and only this maximally
entangled state, violates the inequality (\ref{beta}) is undoubtedly not a
priori predictable. In this context the min-max principle definitely appears
as a powerful tool.

These experiments were carried out in the Quantum Optics Labs at Elsag spa,
Genova, within EC-FET project QAP-2005-015848. The authors thank Caslav
Brukner for helpful discussions.


\end{document}